\documentclass[%
 reprint,
 amsmath,amssymb,
 aps,
]{revtex4-2}
\usepackage{graphicx}
\usepackage{dcolumn}
\usepackage{bm}
\usepackage{braket}
\usepackage{float}
\usepackage{multirow}
\allowdisplaybreaks[1]

\def\leqn#1{(\ref{#1})}

\def\beq{\begin{equation}}
\def\eeq#1{\label{#1}\end{equation}}
\def\eeqn{\end{equation}}

\newenvironment{Eqnarray}%
   {\arraycolsep 0.14em\begin{eqnarray}}{\end{eqnarray}}
\def\beqa{\begin{Eqnarray}}
\def\eeqa#1{\label{#1}\end{Eqnarray}}
\def\eeqan{\end{Eqnarray}}

\def\leqn#1{(\ref{#1})}

\def\ee{e^+e^-}

\begin{document}

\title{The Sensitivity of Higgs Factories to the \\ Minimal Supersymmetric Standard Model via Precision Measurements }

\author{Kamal Maayergi, Devin G.~E.~Walker, Michael Fagundes-Rocha}
\affiliation{Department of Physics and Astronomy, Dartmouth College \\Hanover, NH 03755 USA}%

\author{Michael E.~Peskin}
\affiliation{
SLAC National Accelerator Laboratory\\ Stanford University\\ Menlo Park, California 94025 USA
}%

\begin{abstract}
Precision measurements of Higgs boson couplings at future Higgs
factories provide an indirect probe of new particles at mass scales
beyond the reach of the high luminosity run of the Large Hadron
Collider (HL-LHC). We study this sensitivity in the Minimal
Supersymmetric Standard Model (MSSM), focusing on radiative
corrections to the third-generation Yukawa couplings. In particular, supersymmetric threshold corrections to the bottom Yukawa coupling and, in a smaller parameter region, to the tau Yukawa coupling, can remain significant even for multi-TeV superpartner masses.  We perform a
numerical survey of the MSSM parameter space, using a Markov Chain
Monte Carlo, in order to identify models consistent with a 125 GeV
Higgs boson and exhibiting  potentially observable
Higgs-coupling deviations.  We find that precision measurements of the
bottom Yukawa coupling can probe heavy Higgs masses  of order 2-3 TeV
and retain sensitivity to gluinos, Higgsinos, and bottom squarks with
masses above 5 TeV.  The corresponding sensitivity in the $\tau$
Yukawa coupling is also present but more limited in parameter space.
These results illustrate the characteristic pattern of Higgs-coupling deviations
predicted by the MSSM.  They demonstrate that
precision Higgs measurements can provide sensitivity to supersymmetric
physics well beyond the expected direct-search reach of the HL-LHC.

\end{abstract}

\maketitle
\section{\label{sec:Intro} Introduction}

This is the third in a series of papers
exploring the ability of
measurements of the Higgs boson couplings to access signals
of new physics beyond the Standard Model.   The particle physics
community  envisions the next major collider after the LHC to be a
``Higgs factory'', an $\ee$ collider capable of measuring the mass and
partial widths of the 125 GeV Higgs boson to high
precision~\cite{EuropeanStrategyGroup:2020pow,Endo:2022imj,P5:2023wyd,EuropeanStrategyGroup:2025}.
To set goals for the accelerator and detector designs of a Higgs
factory, it is important to understand how the proposed capabilities
of this facility map onto the expectations for observable new physics
effects
in a wide variety of
beyond-Standard-Model (BSM) theories. In our previous papers, we have explored
this question through detailed scans of the parameter space of
explicit proposed theories -- 
2-Higgs-doublet models (2HDM)~\cite{Maayergi:2025ybi} and 
Little Higgs models illustrating Higgs
compositeness~\cite{Maayergi:2026nbt}.
In this paper, we explore this question for  the Minimal
Supersymmetric Standard Model (MSSM). 

The  opportunities offered by the precision measurement of Higgs
couplings are an important target for future searches for physics
beyond the Standard Model.  Given a precisely known value of the Higgs
boson mass, the Standard Model predictions for these couplings can be
developed to similar precision.   The phenomenological  inputs to these
predictions can be obtained from precision electroweak measurements
and, for QCD inputs, from lattice gauge theory.  The measurements 
depend very weakly on non-perturbative and hadronization effects.  In
contrast to searches for BSM effects in flavor observables, they do
not depend on the scheme of flavor mixing but only on the masses of
predicted new particles and their  order-1 couplings to particles of the Standard
Model.

In the recent literature,  the magnitude of effects on Higgs
couplings from BSM physics
is estimated using effective field
theory~\cite{Dawson:2022zbb,deBlas:2025gyz}.
This approach, while giving important insight, offers only a generic
picture and does not allow estimates of the mass reach of
measurements. That is why it is important to explore the parameter
space of explicit BSM models in detail.   This exploration can display
opportunities that are not apparent from generic estimates.   Our
previous papers showed sensitivity to new particles with masses of
5~TeV and above, well beyond the expected direct discovery reach of
the High-Luminosity LHC.

In this paper, we turn our attention to supersymmetry (SUSY).   Here, a
detailed scan of the parameter space was done  more than 10 years ago
by Hewett, Rizzo, and their students
using the 19-parameter ``phenomenogical MSSM''
(pMSSM)~\cite{Conley:2010du,Cahill-Rowley:2012ydr,Cahill-Rowley:2014wba}.
This scan was based on the public codes
SOFTSUSY~\cite{Allanach:2001kg}  and HDECAY~\cite{Djouadi:1997yw}.
These codes were written at a time when it was not considered
necessary to achieve high accuracy in the decoupling limit with SUSY states in the
multi-TeV mass region. This cast some doubt on the conclusions of the
pMSSM study for models with SUSY states of high mass.
A study done shortly afterward by Endo, Moroi,
and Nojiri, using new formulae for the loop diagrams at high mass,
confirmed the presence of models with multi-TeV squark masses giving
large deviations in the Higgs couplings~\cite{Endo:2015}.   This group
identified the physics that leads to enhanced Higgs couplings to $b$
and $\tau$ in some regions of parameter space, but they did not carry
out a systematic scan over parameter space.
Since the time of that study, the formulae for
SUSY particle masses and the Higgs boson mass have been considerably improved
for models with very heavy SUSY states, adding multi-loop corrections
and careful attention to decoupling, leading up to public codes such
as  SARAH~\cite{Staub:2013tta},  included in the contemporary
framework FlexibleSUSY~\cite{Athron:2017fvs}.
The public codes FeynHiggs~\cite{Bahl:2018qog} and
SUSY-HIT~\cite{Djouadi:2006bz}
provide updated computations of the Higgs  boson couplings and partial widths.\par
There has remained some doubt that precision Higgs coupling
measurements can be sensitive to high-mass SUSY particles.  SUSY
models necessarily contain an extended Higgs sector, but  tree-level
corrections from the extended Higgs Lagrangian lead to rather small
corrections to the Higgs couplings.  The reason for this is
well-known: The leading correction in the decoupling limit comes from
a Lagrangian term that, in SUSY, is a D term in the potential and so
its coupling cannot be larger than a Standard Model gauge
coupling~(see, e.g., \cite{Peskin:2022}).   Thus, large corrections
must come from loop diagrams.   These can be enhanced by large
$\tan\beta$ and large third-generation $A$ terms.  It is somewhat
tricky how decoupling works in this region. A credible analysis must
manifestly illustrate decoupling up to the largest masses of SUSY
particles considered in the study.\par The outline of this paper is as follows.   In Section II, we explain
the main effects that can lead to large BSM contributions to the $b$
and $\tau$ Yukawa couplings, as illustrated by their 1-loop
contributions. We discuss with some care how decoupling is
 satisfied in these calculations.   In Section III, we describe our scan of a 15-parameter
space within the MSSM, using Markov Chain Monte Carlo (MCMC) to
explore the regions of high SUSY particle mass.  We used FlexibleSUSY
to generate the SUSY
spectrum and SUSY-HIT~\cite{Djouadi:2006bz}  to generate the Higgs
couplings.  We discuss our choice of these codes and give
comparisons to other choices.  In Section IV, we describe the
results of the scan.  We find a mass reach that is not as impressive
as in our earlier model explorations.  But still it is substantial,
with sensitivity to models with $m_A \sim 2-3$~TeV, beyond the
expected direct search reach of the HL-LHC, and even higher
masses of the charged and colored SUSY partners.  These results may be
compared to those of the earlier surveys by  Bahl et
al.~\cite{Bahl:2020kwe}  and  the GAMBIT
collaboration~\cite{Athron:2022uzz}.   We consider here higher values
of the SUSY partner masses, and we attempt a complete survey of the
MSSM parameter space rather than restricting our study to specific
scenarios.

\section{Minimal Supersymmetric Standard Model}
\label{sec:composite models}

\newcommand{\sbet}{s_\beta}
\newcommand{\cbet}{c_\beta}
\newcommand{\tbet}{t_\beta}
\newcommand{\salpha}{s_\alpha}
\newcommand{\calpha}{c_\alpha}
\newcommand{\sba}{s_{\beta-\alpha}}
\newcommand{\cba}{c_{\beta-\alpha}}

\subsection{MSSM Higgs Sector and Higgs Coupling Corrections}
\label{subsec:mssm_higgs_background}

In this section, we review the dominant contributions to the shifts in
the Higgs boson couplings, presenting the formulae at the 1-loop
level.   This will give some orientation to the general survey of
these corrections that we will discuss in the following sections.

 The first ingredient in these corrections comes from the
 two-Higgs-doublet structure of the Higgs sector of the MSSM. This
 model contains two 
electroweak scalar doublets, $H_u$ and $H_d$. The 
constraints of supersymmetry require  that the up-type quarks obtain
their massses only from $H_u$ and the down-type quarks obtain their
masses only from $H_d$.  However, this correlation is broken after
supersymmetry breaking.   The corrections that we will discuss make
use of this fact.

The neutral components of $H_u$ and $H_d$ acquire vacuum expectation
values (vevs)
\begin{align}
    \langle H_u^0\rangle
    &=
    \frac{v\,\sin\beta}{\sqrt{2}},
    &
    \langle H_d^0\rangle
    &=
    \frac{v\, \cos\beta}{\sqrt{2}},
    \label{eq:higgs_vevs}
\end{align}
where the angle $\beta$ is defined by the ratio of the vevs
\begin{align} 
\tan\beta
    &\equiv
    \frac{\langle H_u^0\rangle}
    {\langle H_d^0\rangle},
    &
    v
    &\simeq
    246~{\rm GeV}.
    \label{eq:tanbeta_def}
\end{align}
Assuming CP conservation in the Higgs potential, the
mass eigenstates are two neutral CP-even Higgs bosons, $h$ (lighter)
and $H$ (heavier), one CP-odd pseudoscalar $A$, and a charged pair
$H^\pm$.    The lighter neutral Higgs boson $h$ is taken to be the
observed SM-like Higgs boson.
We will mainly consider parameters in the decoupling regime, in which 
$m_A\gg m_Z$  while $H$, $A$, and $H^\pm$ are much
heavier and approximately degenerate in
mass.

The neutral CP-even Higgs fields are related to the mass eigenstates
by
the mixing angle $\alpha$:
\begin{align}
    {\rm Re}(H_u^0)
    &=
    \frac{1}{\sqrt{2}}
    \left(
    v\sin\beta
    +
    h\cos\alpha
    +
    H\sin\alpha
    \right),
   \nonumber  \\
    {\rm Re}(H_d^0)
    &=
    \frac{1}{\sqrt{2}}
    \left(
    v\cos\beta
    -
    h\sin\alpha
    +
    H\cos\alpha
    \right).
    \label{eq:Hud_mixing}
\end{align}
The mismatch between $\alpha$ and $\beta$ gives rise to differences
between the measured quark and lepton masses, which depend on the
Higgs vevs, and the Yukawa couplings, which depend on the structure of
the Higgs mass eigenstates.
The decoupling limit corresponds to $\cos(\beta-\alpha)\to 0$, or, equivalently,
$\sin(\beta-\alpha)\to 1$.   The cosine falls off as $1/m_A^2$ as the
$A$ mass is increased.

In the MSSM, the dominant corrections to the Yukawa couplings come from radiative
corrections due to superparticles that mix the relations between the
quark and lepton types and the Higgs vevs. For the third-generation
quarks, these corrections modify the Higgs coupling Lagrangian to the
form~\cite{Hall:1993gn,Hempfling:1993kv,Carena:1994bv,Carena:1999py,Endo:2015,Anandakrishnan:2014}
\begin{align}
-\mathcal{L}_{\rm eff}
&=
y_b\,\epsilon_{ij}\,
\bar b_R H_d^i Q_L^j
+
\Delta y_b\,
\bar b_R Q_L^k H_u^{k*}
\nonumber\\
&\quad
+
y_t\,\epsilon_{ij}\,
\bar t_R Q_L^i H_u^j
+
\Delta y_t\,
\bar t_R Q_L^k H_d^{k*}
+
{\rm h.c.}
\label{eq:effective_yukawa_lagrangian}
\end{align}
Here $Q_L$ is the third-generation quark doublet, while $t_R$ and $b_R$ are the
right-handed top and bottom fields. The loop-induced ``non-holomorphic''
corrections $\Delta y_b$ and $\Delta y_t$  are finite but
they are not decoupling, in the
sense that they do not fall off as $1/M^2$ when the superparticle mass
parameters $M$ become large. 

The parameters  $\Delta y_b$ and $\Delta y_t$ renormalize the physical
$b$ and $t$ quark masses, according to 
\begin{align}
    m_b
    &=
    \frac{y_b}{\sqrt{2}}
    v\cos\beta
    \left(
    1+
    \frac{\Delta y_b}{y_b}\tan\beta
    \right)
    \nonumber\\
    &\equiv
    \frac{y_b}{\sqrt{2}}
    v\cos\beta
    (1+\Delta_b),
    \label{eq:bottom_mass_delta_b}
    \\
    m_t
    &=
    \frac{y_t}{\sqrt{2}}
    v\sin\beta
    \left(
    1+
    \frac{\Delta y_t}{y_t}\cot\beta
    \right)
    \nonumber\\
    &\equiv
    \frac{y_t}{\sqrt{2}}
    v\sin\beta
    (1+\Delta_t).
    \label{eq:top_mass_delta_t}
\end{align}
where~\cite{Endo:2015,Carena:1999py,Noth:2008tw,Hofer:2009xb,Anandakrishnan:2014}
\begin{align}
    \Delta_b
    &\simeq
    \bigg[
    \frac{2\alpha_s}{3\pi}
    M_3\mu\,
    I(m_{\tilde b_1}^2,m_{\tilde b_2}^2,M_3^2)
    \nonumber\\
    &\hspace{0.65cm}
    +
    \frac{y_t^2}{16\pi^2}
    \mu A_t\,
    I(m_{\tilde t_1}^2,m_{\tilde t_2}^2,\mu^2)
    \bigg]\tan\beta ,
    \label{eq:delta_b}
\end{align}
and
\begin{align}
    \Delta_t
    &\simeq
    \bigg[
    \frac{2\alpha_s}{3\pi}
    M_3\mu\,
    I(m_{\tilde t_1}^2,m_{\tilde t_2}^2,M_3^2)
    \nonumber\\
    &\hspace{0.65cm}
    +
    \frac{y_b^2}{16\pi^2}
    \mu A_b\,
    I(m_{\tilde b_1}^2,m_{\tilde b_2}^2,\mu^2)
    \bigg]\cot\beta .
    \label{eq:delta_t}
\end{align}
Here $M_3$ is the gluino mass, $\mu$ is the Higgsino mass parameter,
and   $A_b$, $A_t$ are the soft
supersymmetry breaking $A$ terms. The loop
function is \cite{Endo:2015,Anandakrishnan:2014}
\begin{align}
    I(a,b,c)
    =
    \frac{
    ab\ln(a/b)
    +
    bc\ln(b/c)
    +
    ca\ln(c/a)
    }
    {(a-b)(b-c)(a-c)} .
    \label{eq:loop_integral_I}
\end{align}
For fixed squark masses, of order 1~TeV, and  large
gluino mass, the leading term in $\Delta_b$ is proportional to
$\alpha_s (\mu/M_3)\log M_3$, which can remain large for large values
of the mass parameters.   Therefore, the down-type Higgs couplings can receive
appreciable corrections even when the superpartner masses are heavy.
The correction $\Delta_b$ is enhanced at large $\tan\beta$, while $\Delta_t$ is
suppressed by $\cot\beta$.

 The term $\Delta_b$ can lead to an order-1 shift in the $b$ quark
 mass.
 This point was especially emphasized by Anandakrishnan and
 collaborators  \cite{Anandakrishnan:2014}, who were interested in the
 implications for models of  Yukawa unification.  In the more general
 context of the MSSM, this effect is not directly observable, since
 the renormalization of the MSSM
requires the values of the fermion masses as inputs.

To find the observable effect of a discrepancy between the Higgs boson
coupling to $b\bar b$ and the Standard Model prediction, we need to
combine the above result for $\Delta_b$ with the expression for the 
Higgs mass eigenstates in \leqn{eq:Hud_mixing}.
At 1 loop, the coupling of $b\bar b$ to the light Higgs mass
eigenstate 
\begin{equation}
    -\mathcal{L}_{hf\bar f}
    \equiv
    g_{hf\bar f}\,h\bar f f .
    \label{eq:hff_coupling_definition}
\end{equation}
can be seen to be \cite{Endo:2015}
\begin{align}
    g_{h\bar b b}
    &=
    -
    \left(
    \frac{\sin\alpha}{\cos\beta}
    \right)
    \frac{
    1-\Delta_b\cot\alpha\cot\beta
    }
    {1+\Delta_b}
    g_{h\bar b b}^{\rm SM}
    \nonumber\\
    &=
    \bigg[
    \sin(\beta-\alpha)
    -
    \frac{
    \tan\beta-\Delta_b\cot\beta
    }
    {1+\Delta_b}
    \cos(\beta-\alpha)
    \bigg]
    g_{h\bar b b}^{\rm SM}.
    \label{eq:hbb_coupling}
\end{align}
Equivalently, defining the coupling modifier
\begin{equation}
    \kappa_b
    \equiv
    {g_{h\bar b b}} / {g_{h\bar b b}^{\rm SM}},
\end{equation}
this can be written 
\begin{align}
    \kappa_b
    &=
    \sin(\beta-\alpha)
    -
    \frac{
    \tan\beta-\Delta_b\cot\beta
    }
    {1+\Delta_b}
    \cos(\beta-\alpha).
    \label{eq:kappa_b}
\end{align}

The formula \leqn{eq:kappa_b} clarifies the  physics that leads to
larger than expected  deviations of the Higgs boson coupling to $b$ from the Standard
Model expectation.   We have noted already that the non-holomorphic
correction $\Delta_b$ is non-decoupling, but that this is allowed,
because it contributes to the renormalization of the $b$ quark mass.
We have also noted that the shift of the $b$ quark mass can be an
order-1 effect. $\Delta_b$, and this mass shift, can be of either
sign.
Now we see also that the effect of the coupling
modification is multiplied by $\cos(\beta-\alpha)$, which goes to 0
in the decoupling limit.  Thus, decoupling of the heavy particle
effects is properly satisfied.  However, since $|\Delta_b|$ can be
large,  even when $\sin(\beta-\alpha)$ is very close to 1, the term proportional to
$\cos(\beta-\alpha)$ can cause a measurable deviation in the bottom
Yukawa coupling.  The effect can be enhanced at large $\tan\beta$.  In
this limit, the main effect comes from the factor $1/(1+\Delta_b)$,
which is exactly the inverse of the $b$ mass shift due to the non-holomorphic
corrections.

This has an implication for the region of parameter space in which the
effect on $\kappa_b$ is largest.   The value of $m_A$ must not be too
large, since this sets the suppression from decoupling.  On the other
hand, the effect on $\kappa_b$ can be seen for very large values of
the gluino, higgsino, and squark masses, well above 5~TeV. 

The $\tau$ Yukawa coupling is shifted in  a similar way.   For $\tau$,
there is no gluino correction, but there can be a significant
contribution from wino-Higgsino loops~\cite{Endo:2015}.
\begin{align}
    \kappa_\tau
    &=
    \sin(\beta-\alpha)
    -
    \frac{
    \tan\beta-\Delta_\tau\cot\beta
    }
    {1+\Delta_\tau}
    \cos(\beta-\alpha).
    \label{eq:kappa_tau}
\end{align}
The relevant loop correction is \cite{Endo:2015}
\begin{align}
    \Delta_\tau
    &\simeq
    -
    \frac{3\alpha_2}{8\pi}
    M_2\mu\tan\beta\,
    I(m_{\tilde\tau_L}^2,M_2^2,\mu^2),
    \label{eq:delta_tau}
\end{align}
Here,  $M_2$ is the wino mass, and the effect is proportional to the weak
$SU(2)$ coupling $\alpha_2$ rather than  $\alpha_s$.

For the top quark, the formula for the corrected Higgs coupling is
\begin{align}
    g_{h\bar t t}
    &=
    \left(
    \frac{\cos\alpha}{\sin\beta}
    \right)
    \frac{
    1-\Delta_t\tan\alpha\tan\beta
    }
    {1+\Delta_t}
    g_{h\bar t t}^{\rm SM}
    \nonumber\\
    &=
    \bigg[
    \sin(\beta-\alpha)
    +
    \frac{
    \cot\beta-\Delta_t\tan\beta
    }
    {1+\Delta_t}
    \cos(\beta-\alpha)
    \bigg]
    g_{h\bar t t}^{\rm SM}.
    \label{eq:htt_coupling}
\end{align}
Thus,
\begin{align}
    \kappa_t
    \equiv
    \frac{g_{h\bar t t}}
    {g_{h\bar t t}^{\rm SM}}
    &=
    \sin(\beta-\alpha)
    +
    \frac{
    \cot\beta-\Delta_t\tan\beta
    }
    {1+\Delta_t}
    \cos(\beta-\alpha).
    \label{eq:kappa_t}
\end{align}
Unlike the bottom Yukawa coupling, the leading correction to the top
coupling
is not enhanced by $\tan\beta$, since $\Delta_t$ itself is proportional to
$\cot\beta$.    Unfortunately, this coupling can only be measured with
precision at
$\ee$ colliders with energies above the top quark threshold, and the ultimate
expected precision is lower, about 1\%.

In principle, the Higgs boson couplings to $W$, $Z$, $g$, and $\gamma$
should also show corrections induced by loop effects of SUSY
particles.  However, we did not find a region of the MSSM parameter
space in which these corrections are large enough to be observed given
the currently estimated capabilities of Higgs factories.   This
feature distinguishes the signatures of SUSY and of extended Higgs sectors
from the signatures of models with Higgs compositeness,
illustrated in \cite{Maayergi:2026nbt}.

\section{Numerical survey of Higgs-Yukawa deviations}
\label{sec:numerical_survey}

Having described the origin of the Higgs-Yukawa modifications in
Sec.~\ref{subsec:mssm_higgs_background}, we now
discuss the survey of the MSSM parameter
space.  The goal of the scan is to efficiently populate regions of
parameter space in which the light Higgs mass is close to \(125~{\rm
  GeV}\) and the Yukawa coupling deviations from the
SM may be significant and observable at Higgs factories.
In particular, we are interested in whether percent-level deviations
in the bottom and tau Yukawa couplings can persist when the heavy
Higgs bosons and the
superpartners are in the multi-TeV regime.

\begin{table}[t]
\centering
\renewcommand{\arraystretch}{1.25}
\begin{tabular}{c c}
\hline\hline
Parameter or quantity & Scan Range \\
\hline
$m_A$ & $m_A \geq 1000~{\rm GeV}$ \\
$m_h$ & $\simeq 125~{\rm GeV}$ \\
$\tan\beta$ & $\tan\beta \leq 50$ \\
$|\mu|$ & $|\mu| \leq 10000~{\rm GeV}$ \\
$|M_1|$ & $|M_1| \leq 3000~{\rm GeV}$ \\
$|M_2|$ & $|M_2| \leq 3000~{\rm GeV}$ \\
$|M_3|$ & $|M_3| \leq 8000~{\rm GeV}$ \\
$m_{Q_3},\,m_{U_3},\,m_{D_3}$ & $0 < m_{Q_3},m_{U_3},m_{D_3} \leq 10000~{\rm GeV}$ \\
$m_{L_3},\,m_{E_3}$ & $0 < m_{L_3},m_{E_3} \leq 8000~{\rm GeV}$ \\
$|A_t|$ & $|A_t| \leq 12000~{\rm GeV}$ \\
$|A_b|$ & $|A_b| \leq 12000~{\rm GeV}$ \\
$|A_\tau|$ & $|A_\tau| \leq 10000~{\rm GeV}$ \\
\hline\hline
\end{tabular}
\caption{MSSM parameters varied in our scan of MSSM parameter space.}
\label{tab:moderate-input-ranges}
\end{table}

\begin{figure*}
\includegraphics[width=.7\linewidth]{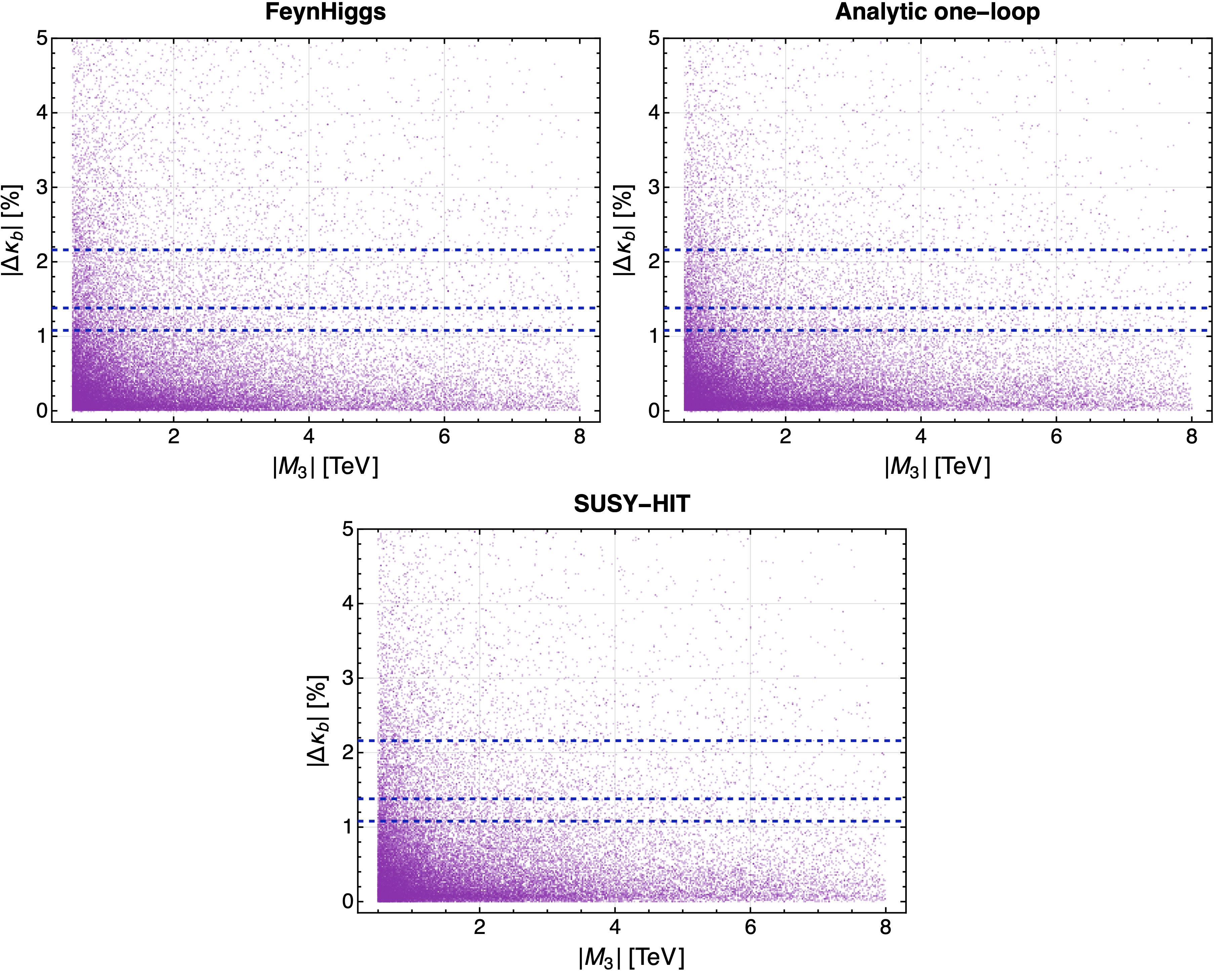}
\caption{Comparision of the evaluations of $\Delta\kappa_b$ as a
  function of the gluino mass parameters by FeynHiggs 2.18.1, SUSY-HIT 1.5a, and direct evaluation of the formulae
  given in Section II.}
\label{fig:kbthree}
\end{figure*}

We aim for a thorough exploration of the effects of the MSSM on
third-generation Yukawa couplings.  With this in mind, we scanned a
space of 15 parameters, listed in
Table~\ref{tab:moderate-input-ranges}.  These include the Higgs sector
parameters $m_A$ and $\tan\beta$, the $\mu$ parameter and the
three gaugino masses, and the third-generation squark and slepton
masses and $A$ parameters.

To derive the corrections to the $b$ and $\tau$ Yukawa couplings, we
first coded each parameter set into SLHA format for input to the
public code FlexibleSUSY 2.9.0~\cite{Athron:2017fvs}.   This code
generates the
full SUSY
spectrum and consistent coupling constants.  It includes higher-loop
corrections to the mass spectrum, including results for the Higgs
boson mass up to 3 loops~\cite{Slavich:2020zjv}.

In performing the scan over our parameter space, an issue is that the
Higgs boson mass is an output of  FlexibleSUSY and will not equal the
observed Higgs mass for general inputs.  To address this problem, we
explored the parameter space using a 
Metropolis-Hastings Markov chain Monte Carlo
algorithm \cite{Speagle:2019ffr}. The MCMC chains were constructed
to minimize the difference between the computed Higgs boson mass and
the value of 125~GeV. MCMC can tunnel through barriers to find regions
that are not directly connected.  Since the purpose of the scan was
exploratory coverage, the accepted points should be interpreted as a
survey of viable and phenomenologically interesting
regions.  In our plots, the density of points shown has no physical
significance.   Chosen points were accumulated in files, with their
SLHA files saved for the evaluation of $\Delta\kappa_{b,\tau}$.

The numerical scan was  performed over the parameter ranges shown in
Table~\ref{tab:moderate-input-ranges}.  The scan was  designed to cover a
broad multi TeV MSSM parameter space while avoiding regions in which
the threshold expansion becomes numerically delicate. The bounds
chosen were based on the scan performed in \cite{Anandakrishnan:2014},
but were extended more broadly to explore a larger parameter space. We
allowed large values of the higgsino mass parameter, gaugino masses,
third-generation scalar masses, and trilinear couplings in order to
probe the regime relevant for indirect Higgs coupling measurements at
Higgs factories. At the same time, the ranges were chosen to avoid
extremely large hierarchies that could make the largest deviations
depend on isolated corners of parameter space, where the values are
potentially subject to large numerical errors.

  Soft SUSY-breaking masses
  not listed in Table~\ref{tab:moderate-input-ranges}
  were set very large in order to decouple the effects of the
SUSY partners of light quarks and leptons.  The parameter ranges
scanned are listed in the table.  We imposed the lower limit $m_A >
1$~TeV.   We did not impose lower cutoffs on the soft SUSY-breaking
parameters, but our scan naturally avoided the region of light
SUSY particles since, in general, large loop effects  are
necessary within the MSSM to raise the Higgs boson mass to 125~GeV.
For example, our scan found that for
$|\mu| < 800$~GeV the density of points found was much
lower than in the larger $|\mu|$ regions.

Due to the nature of MCMC scans, they tend to populate the most likely
regions of parameter space. This presented as a problem in our case
because our objective was to probe the regions of high coupling
deviation, which are less probable than the points that lie on the
main curves. In order to work around this, the scan was conducted in
three parts, for high, middle and low $m_A$ values, and the MCMC was
also forced to search in regions of interest where high coupling
deviations could be possible to see if there were any valid points in
those areas. The three scans were then stitched together in all of the
upcoming figures. In this way we were better able to ensure that we had
probed and located as many islands of viable points as possible.   The
total number of MSSM points identified by the scan was about 25,000.

 From this starting
point, we tried a number of approaches to produce the Higgs boson
couplings. Using our list of scan points as input, we evaluated the
Higgs boson couplings generated by the codes
FeynHiggs~\cite{Bahl:2018qog} (FeynHiggs 2.18.1) and
SUSY-HIT~\cite{Djouadi:2006bz}  (SUSY-HIT 1.5a). In principle, the  Higgs
boson coupling deviations could be defined in one of two ways.
First, they could be computed directly from the couplings generated by
the codes:
\beq
\Delta \kappa_{b,\tau} = { g(Hff)/ g(Hff)|_{SM}} - 1 \ .
\eeq{firstdeltakappa}
Second, they could be computed from the ratio of partial widths,
\beq
\Delta \kappa_{b,\tau} = [ \Gamma(H\to ff)/ \Gamma(Hff)|_{SM}]^{1/2} - 1 \ .
\eeq{seconddeltakappa}
We will discuss the results with both methods in a moment.   FlexibleSUSY
also
includes a component that computes
Higgs partial widths, called FlexibleDecay~\cite{Athron:2021kve}.
This code does not produce the effects that we have described arising
from 
large values of $\Delta_b$ and $\Delta_\tau$.   This is most likely
the result of defining the Yukawa couplings through the ``decoupling''
renormalization scheme, their eq. (1b).  In their footnote 4, they
note that this prescription does not treat the $\Delta_b$
contributions correctly.   Their renormalization scheme may be the
best for other uses of this general-purpose program, but it omits the
physics that is crucial for our specific application.

For the first method, we found good agreement with the results from
FeynHiggs and SUSY-HIT.  In most regions of the
parameter space, these public  codes agree in the values of $\kappa_b$ and
$\kappa_\tau$ to within  about 0.2\%,  although we have seen 1\%
discrepancies in regions with the largest values of $\Delta_b$.
Deviations seem to be due to the different prescriptions for resumming
large $\tan\beta$ effects.  The
overall pattern of $\kappa_b$ values over the the set of scan points is very
similar for the two evaluations and also agree well with the direct
evaluation of the 1-loop formulae.  
Figure.~\ref{fig:kbthree} compares  the pattern shown in the values of
$\Delta\kappa_b$ as a function of the gluino mass $M_3$ for the
evaluation from the two codes and from the 1-loop expressions.

For the second method, our experience was less satisfying.  The
partial widths
computed by the public codes make use of the Higgs couplings
but add SUSY-QCD and (for FeynHiggs) SUSY-electroweak
corrections.   The $\Delta\kappa$ values are typically larger; however,
they show some conceptual difficulties. Using the
partial widths from FeynHiggs, we found physically sensible results
for $b\bar b$ but clear violation of decoupling at large $m_A$ values
for $\tau^+\tau^-$.  Our results from SUSY-HIT were just the opposite:
large violations of decoupling for $b\bar b$ but not for
$\tau^+\tau^-$.  It is very difficult to
write a public code that answers the needs of all customers.
All corrections included must contain cancellations that implement
decoupling,  These cancellations must
become increasingly exact as the squark, higgsino, and gaugino masses
become large. They can fail for the very large mass values if, for
example, the resummation of large 
$\tan\beta$ terms  is done differently in  the two cancelling expressions.  

 In Section IV, we will show plots of the distribution of values of
 $\Delta \kappa_{b}$ and $\Delta \kappa_{\tau}$ computed using the
 definition \leqn{firstdeltakappa}. We consider this a conservative
 choice.  For definiteness, we plot the
 results obtained from FeynHiggs.    The plots generated from SUSY-HIT
 are very similar.

\begin{figure*}
\includegraphics[width=.8\linewidth]{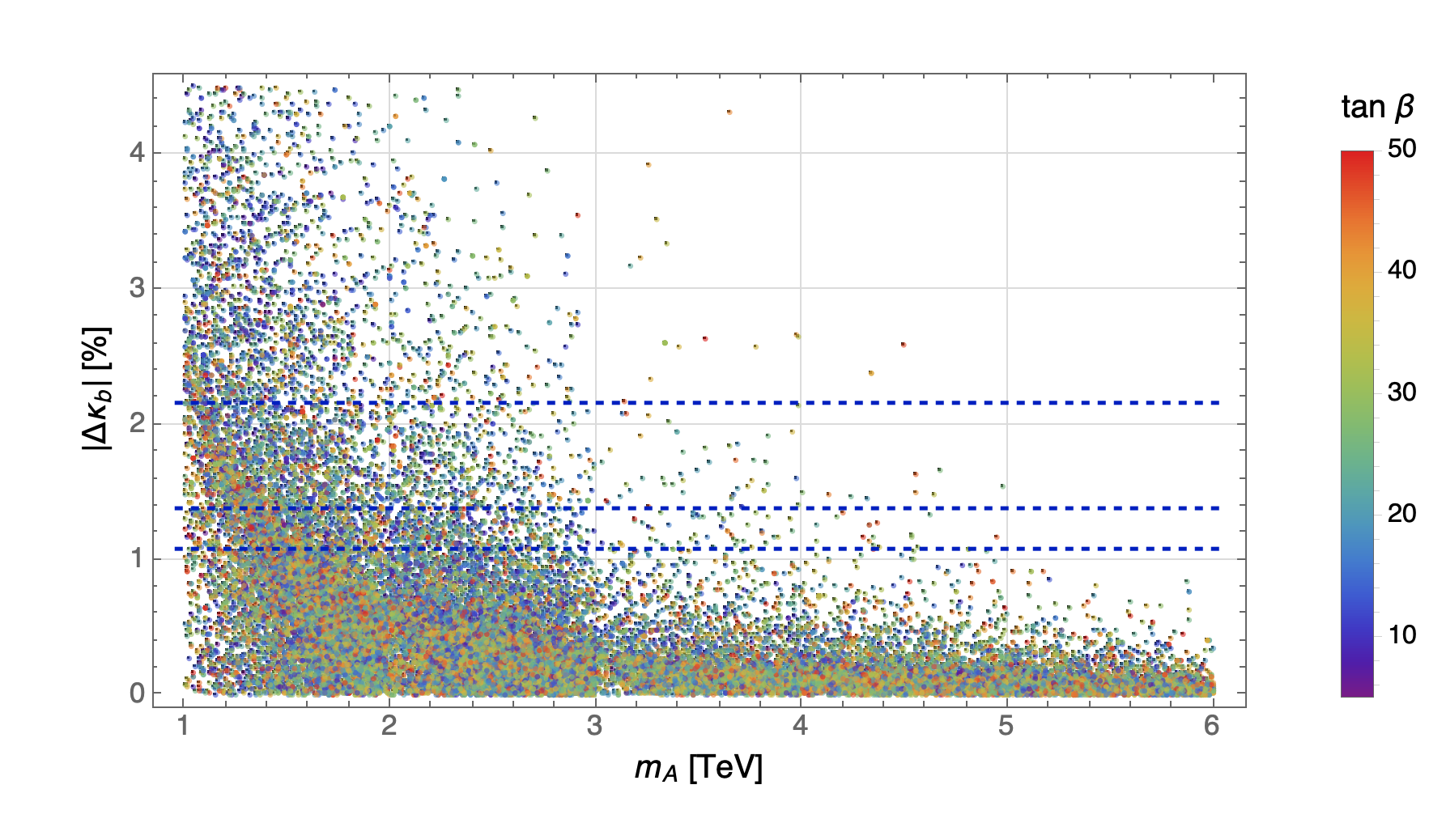}
\caption{Bottom Yukawa coupling deviation as a function of the
  pseudoscalar Higgs mass $m_A$.  Points in this plot are color-coded
  by the value of $\tan\beta$. The largest deviations occur at low
$m_A$, but   percent-level effects can persist into the multi-TeV region
  when the bottom quark  loop corrections $\Delta_b$ enhance the
  effect.  Here and in the following plots, the three horizontal
  dashed lines correspond to 3$\sigma$ sensitivity expected  for three
stages of the LCF Higgs factory proposed for CERN.  The bottom line
also gives the expectation for the FCC program. See the text for more details.}
\label{fig:kb}
\end{figure*}

\begin{figure*}
\includegraphics[width=.8\linewidth]{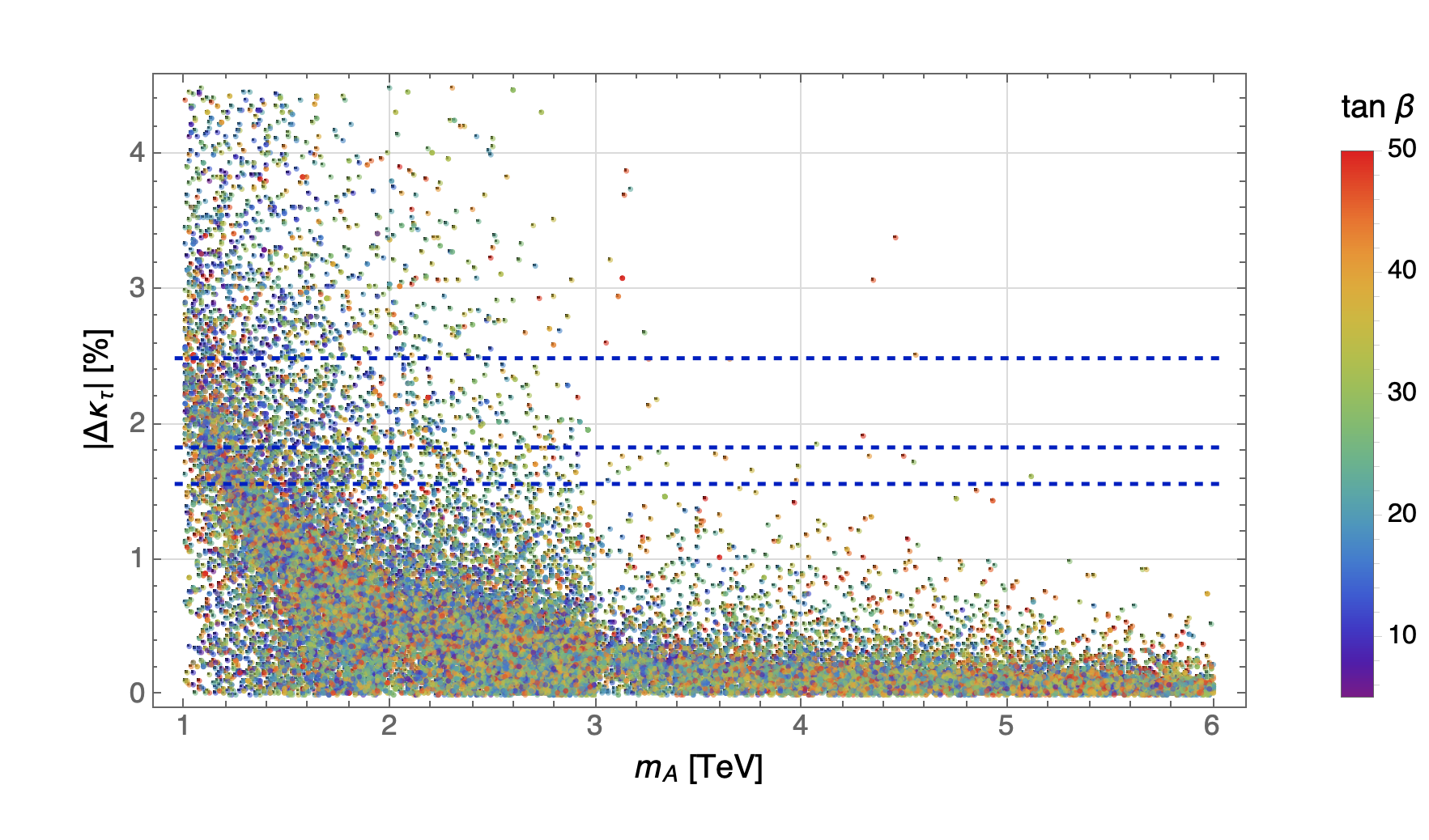}
\caption{Tau Yukawa coupling deviation as a function of the
  pseudoscalar Higgs mass $m_A$.  Points in this plot are color-coded
  by the value of $\tan\beta$.  For tau, there is no gluino-induced
  loop correction, and so the enhancement
  $\Delta\kappa_\tau$ is generally smaller and extends less far in $m_A$.}
\label{fig:ktau}
\end{figure*}

\begin{figure*}
\includegraphics[width=.8\linewidth]{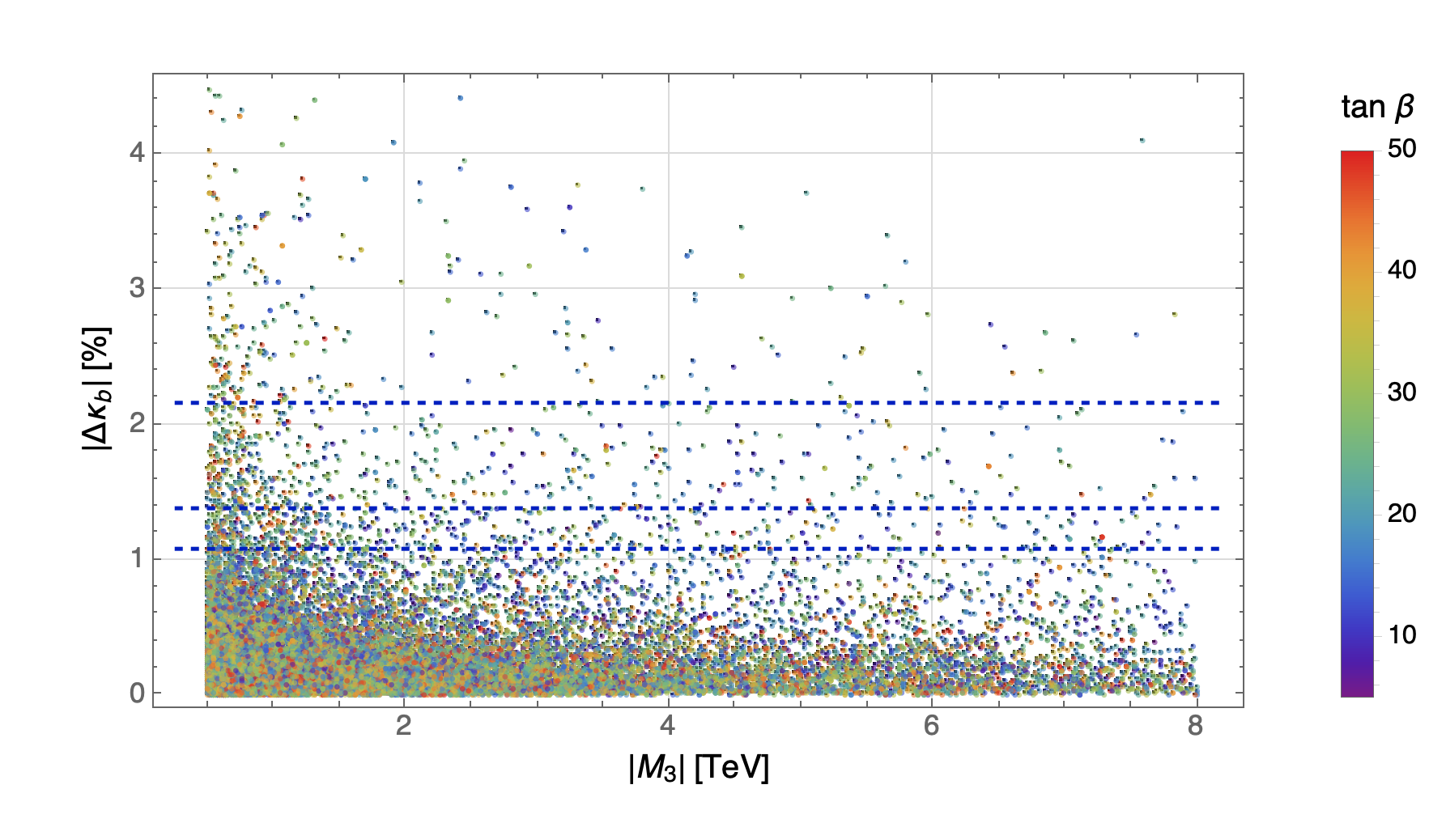}
\caption{Bottom Yukawa coupling deviation as a function of the
  gluino mass parameter $m_3$.  Points in this plot are color-coded
  by the value of $\tan\beta$.   The plotted points are restricted to
  those with  $m_A > 2 \text{ TeV}$.}
\label{fig:kbM3}
\end{figure*}

\begin{figure*}
\includegraphics[width=.8\linewidth]{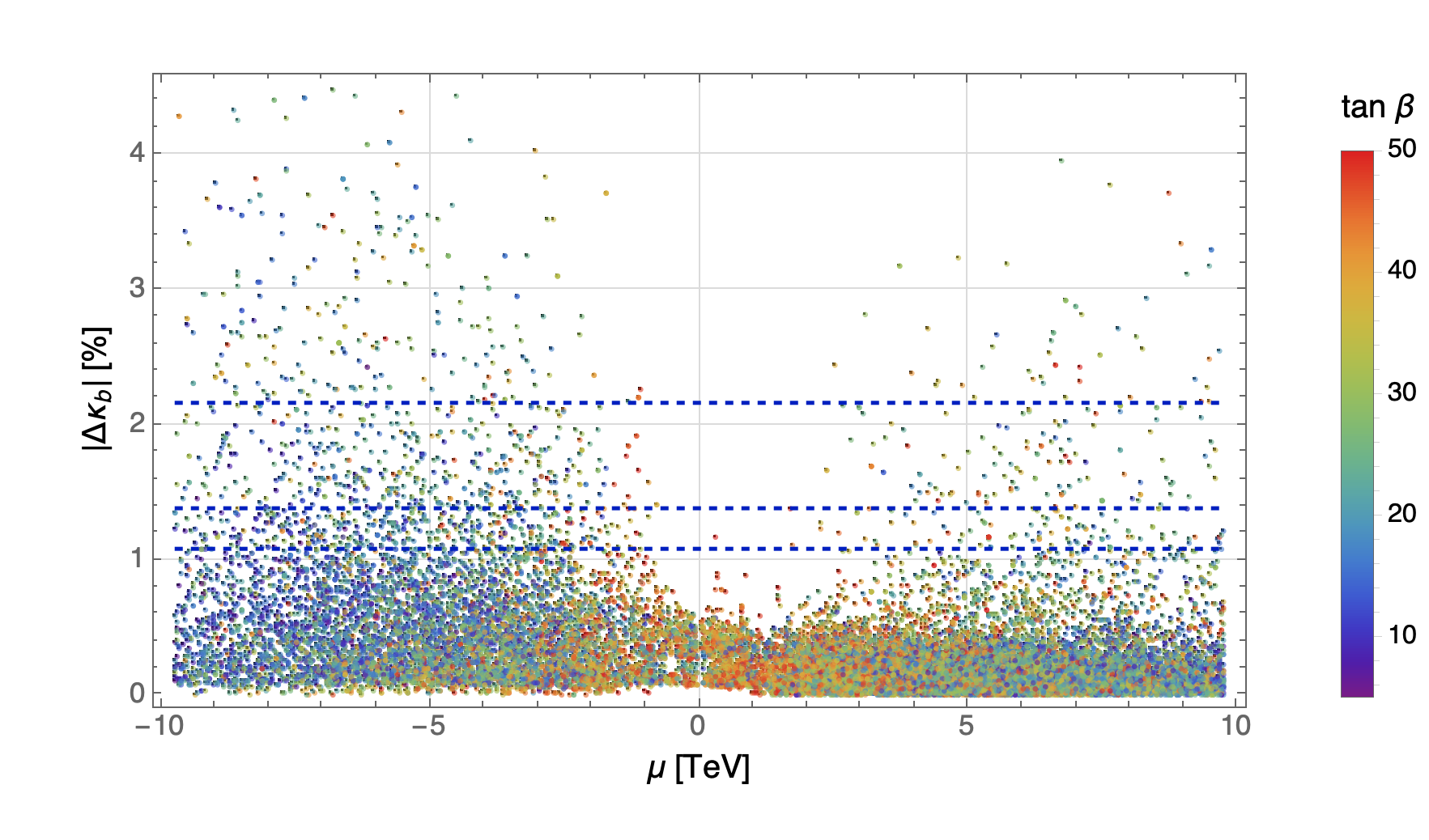}
\caption{Bottom Yukawa coupling deviation as a function of the
  Higgsino mass parameter $\mu$.  Points in this plot are color-coded
  by the value of $\tan\beta$.   The plotted points are restricted to
  those with  $m_A > 2 \text{ TeV}$.}
\label{fig:kbmu}
\end{figure*}

\begin{figure*}
\includegraphics[width=.8\linewidth]{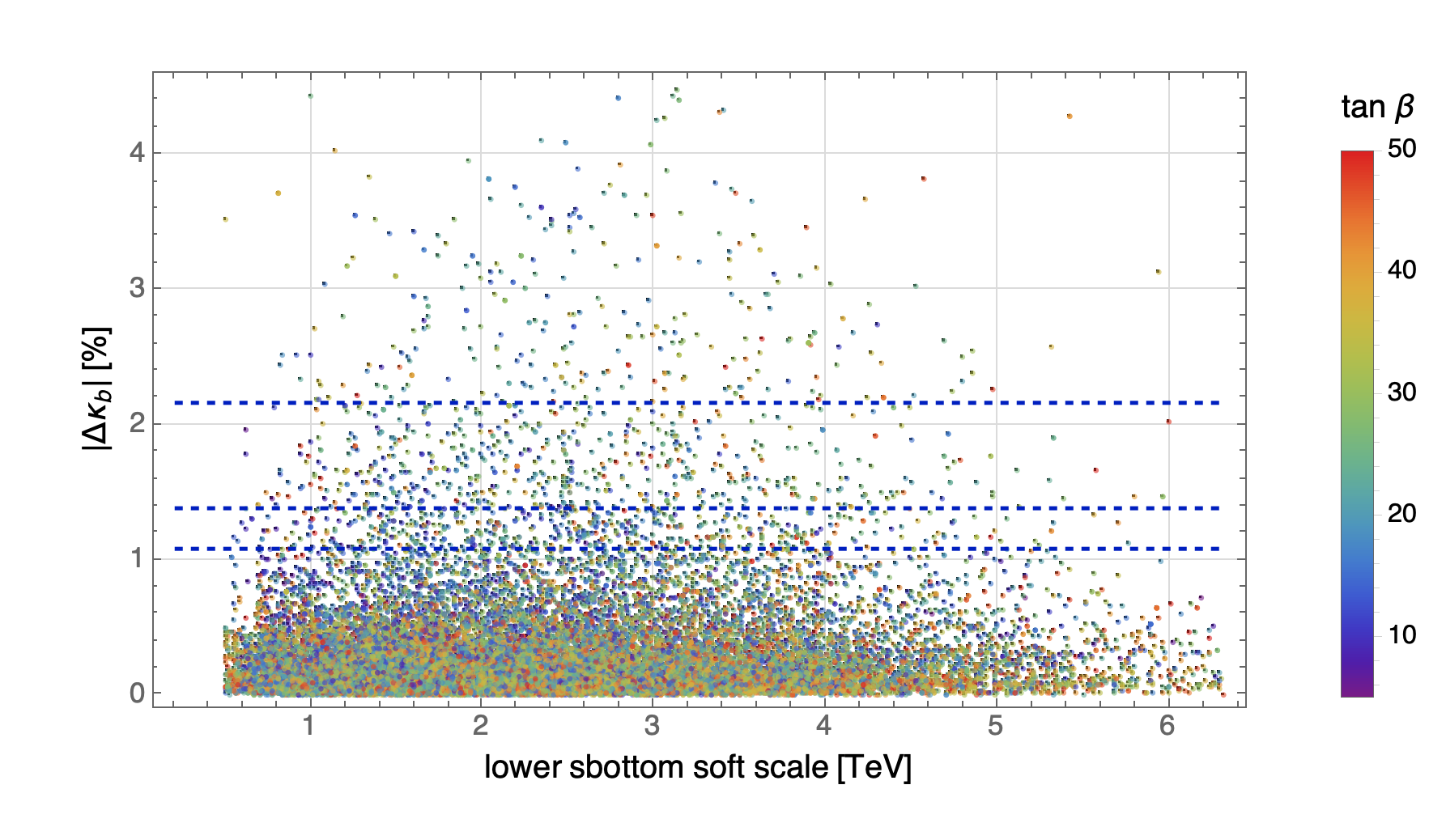}
\caption{Bottom Yukawa coupling deviation as a function of the mass of
  the
  lighter bottom squark.  Points in this plot are color-coded
  by the value of $\tan\beta$.   The plotted points are restricted to
  those with  $m_A > 2 \text{ TeV}$.}
\label{fig:kbsb}
\end{figure*}

\begin{figure*}
\includegraphics[width=.8\linewidth]{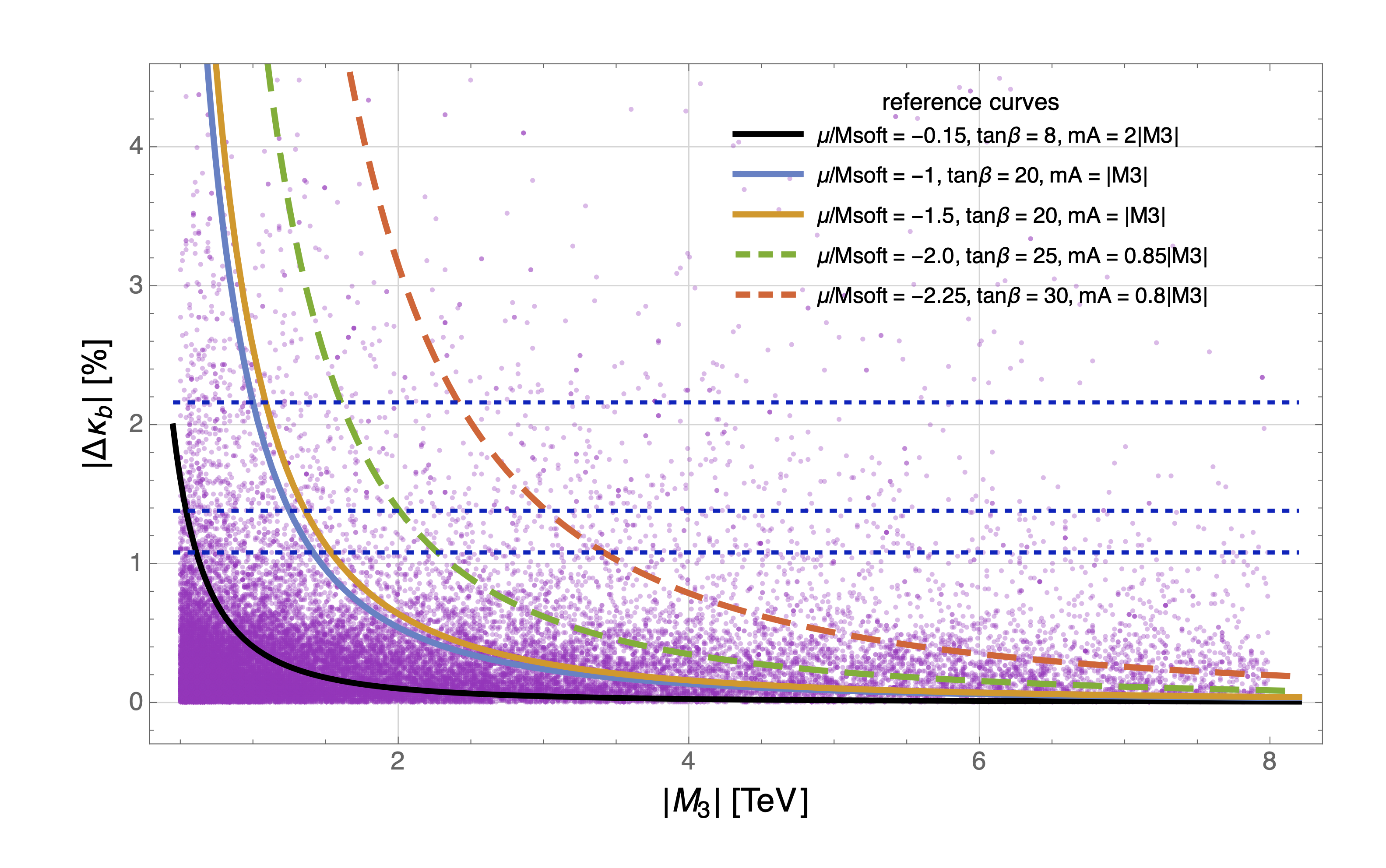}
\caption{Bottom Yukawa coupling deviation as a function of the
  gluino mass parameter $M_3$.  Points in this plot are color-coded
  by the value of $\tan\beta$.   The plotted points are restricted to
  those with  $m_A > 2 \text{ TeV}$.   The lines through the scatter plot show
the 1 loop coupling deviations at illustrative slices through the
parameter space, given in the text.}
\label{fig:kbM3annotated}
\end{figure*}

\begin{figure*}
\includegraphics[width=.8\linewidth]{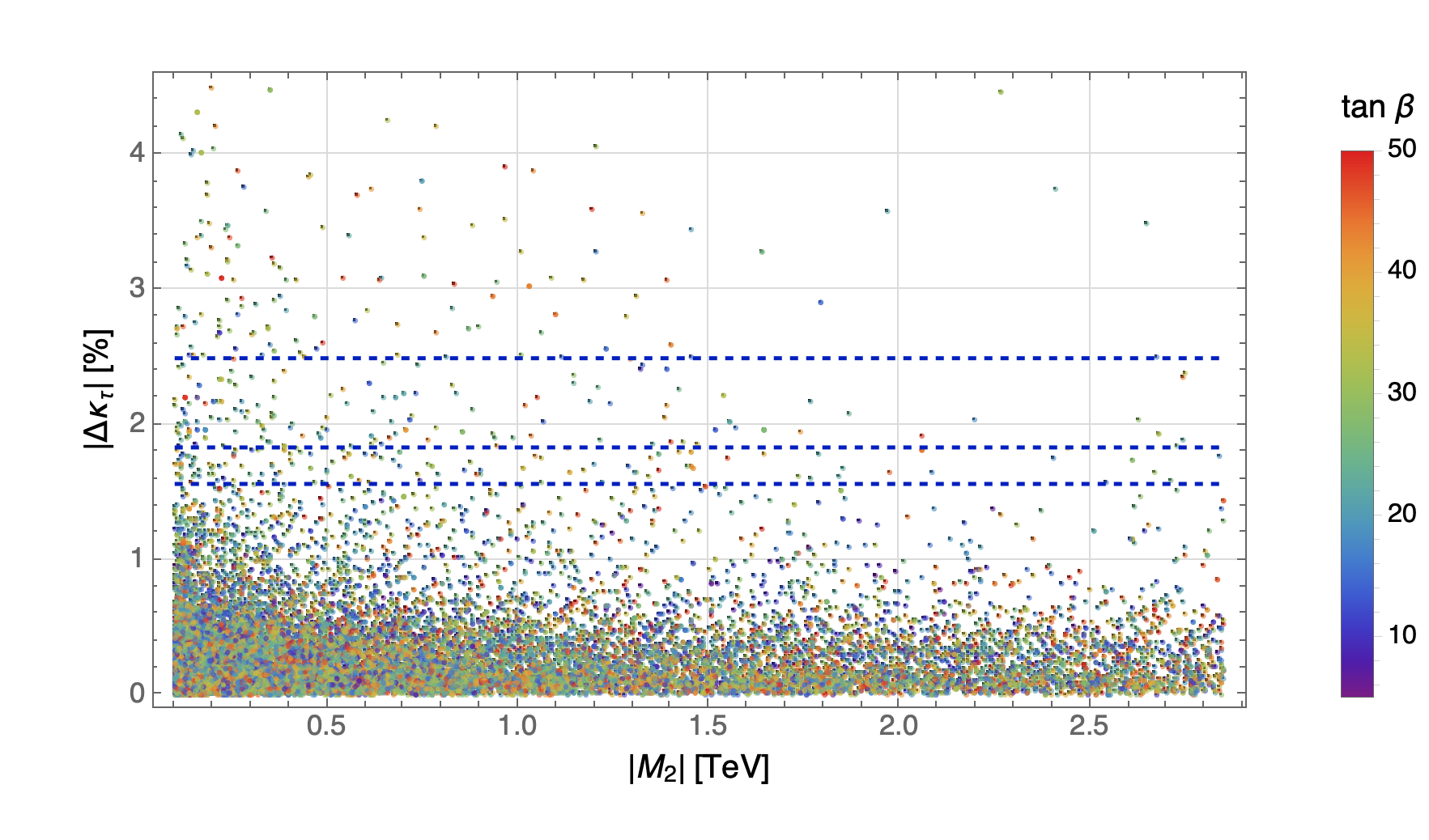}
\caption{Tau Yukawa coupling deviation as a function of the
wino mass parameter $M_2$.  Points in this plot are color-coded
  by the value of $\tan\beta$.   The plotted points are restricted to
  those with  $m_A > 2 \text{ TeV}$. }
\label{fig:ktauM2}
\end{figure*}

For consistency with our earlier papers, we present the scan results
for the deviations of Yukawa couplings  from the Standard Model
predictions compared to the expectations for a 3$\sigma$ deviation at
three stages of the program of the proposed Linear Collider Facility
(LCF) at CERN~\cite{LinearColliderVision:2025hlt}. Usually, a
3$\sigma$ deviation in one partial width leads to a 5$\sigma$
incompatibility with the Standard
Model in a global fit.   The first stage of the LCF
corresponds to 3~ab$^{-1}$ of polarized $\ee$ data at 
250~GeV.  The
lowest line, which includes  8~ab$^{-1}$ of data at 500 or 1000 GeV,
is also the expected result for the FCC-ee program at 240~GeV and
368~GeV~\cite{FCC:2025lpp}.  The expectation for 10~ab$^{-1}$ of data
at 240~GeV alone  is already quite close to this line~\cite{Peskin:2026mhr}.
It has recently been shown that taggers based on Machine Learning have
the capability of dramatically improving flavor tagging at $\ee$
colliders~\cite{Qu:2019gqs,Zhu:2023xpk,Liang:2023wpt}.   Thus, these
expectations are very likely to be superseded in practice.

\section{Results of the survey}

In the previous sections, we have discussed the physics that leads to
observable enhancements in the $Hbb$ and $H\tau\tau$ couplings.
We have also explained
that a systematic survey of the MSSM parameter space in the region of
large SUSY partner masses using current public codes gives a
distribution of Higgs coupling
deviations $\Delta\kappa_b$  that is in general accord with the 1-loop
intuition.   What remains is to display the quantitative results and
the interplay with the sensitivity to $\Delta\kappa_b$ expected at future
Higgs factories.

To begin, Fig.~\ref{fig:kb} shows the dependence of $\Delta\kappa_b$
on the heavy Higgs mass $m_A$.   The decoupling proportional to
$m_A^{-2}$ is clearly visible.  Experimental limits on $m_A$ from the
LHC are 
model-dependent and typically require a simple decay pattern such as
$A, H \to
\tau^+\tau^-$ or $b\bar b$.    Current exclusions from ATLAS and CMS  for
$\tan\beta > 30$ reach $m_A > 1.2$~TeV.    These are expected to extend
to $m_A > 2$~TeV with the HL-LHC data, with stronger exclusions for
$\tan\beta > 30$, if there is no
discovery~\cite{ATLAS:2022hsp}.  Therefore, it is most interesting to
ask for the expectations for observable values of $\Delta\kappa_b$ for
$m_A > 2$~TeV.    We will make this restriction in the following
plots.   If evidence for the heavy Higgs sector is found at the
HL-LHC, then precision measurement of the 125~GeV Higgs couplings
would be even more highly motivated.

Figure~\ref{fig:ktau} shows the dependence of of $\Delta\kappa_\tau$
on the heavy Higgs mass $m_A$.  In this case, without the enhancement
of $\Delta_\tau$ by gluino effects proportional to $\alpha_s$, the
possible deviation cuts off sharply above $m_A = 2$~TeV and hardly
extends beyond 3~TeV.

Figures~\ref{fig:kbM3}, \ref{fig:kbmu}, and \ref{fig:kbsb} show
scatter plots of the values of $|\Delta\kappa_b|$ found in our scan
versus the mass parameters that contribute to $\Delta_b$:  the gluino
mass $M_3$, the lighter bottom squark mass, and  the Higgsino mass
parameter $\mu$, with all points restricted to the region $m_A > 2$
TeV.  Though the typical expectation for $|\Delta\kappa_b|$ is small,
we find a substantial population of points with observable Yukawa
coupling deviations.   As we expected from the 1-loop formulae, the
particle masses associated with these points extend out to very large
values, even to values above 5~TeV.

The plot in terms of $\mu$ is especially instructive.   The large
values of $|\Delta\kappa_b|$ are found in the region of 
\beq
\mu < 0 \ , \ \ M_3 > 0\, \ \ A_t > 0 \ ,
\eeqn
such that $\Delta_b < 0$.    According to \leqn{eq:kappa_b}, the
denominator of $\Delta\kappa_b$ is enhanced for negative values of
$\Delta_b$. These points also typically have $\cos(\beta-\alpha) < 0$, leading to
a positive value of $\Delta\kappa_b$.

A surprising feature of the plots is that points with large
deviations $\Delta\kappa_b$ are distributed over the full range of
$\tan\beta$.  Although the parameter $\Delta_b$ is enhanced at large
$\tan\beta$,  this can be compensated by the  parameter dependence of
the decoupling parameter $\cos(\beta-\alpha)$. 

For further clarity on the origin of large deviations,
Fig.~\ref{fig:kbM3annotated} shows the dependence of $|\Delta\kappa_b|$
on $M_3$ along some illustrative lines in the parameter space.

The comparable dependence of $\Delta\kappa_\tau$ is illustrated in
Fig.~\ref{fig:ktauM2}.  Again, points with relatively large values of
the gaugino mass 
$M _2$ extend to quite high values.  And, again, these points are
found for the full range of values of $\tan\beta$.

\section{Conclusions}

In this paper, we have surveyed the ability of the Minimal
Supersymmetric Standard Model to produce  observable deviations in the
couplings of the Higgs boson as measured with high precision at Higgs
factories.  Although supersymmetry remains a compelling idea for
physics beyond the Standard Model, the many searches for
supersymmetric particles at the LHC have eliminated much of the
low-lying parameter space, forcing us to search for these particles at
masses in the TeV and even multi-TeV region. Searches for
supersymmetric particles will be extended during the period of the
HL-LHC.  But it is important to be aware that the measurements of the
$b$ quark and $\tau$ Yukawa couplings add further ability to search for effects of
supersymmetric particles. 

Unlike the BSM models studied in our earlier
papers~\cite{Maayergi:2025ybi,Maayergi:2026nbt}, we do not find in the
MSSM large and 
robust regions of parameter space in which observable Higgs couplings
appear for multi-TeV masses.   This is because supersymmetry generally
does not allow large dimensionless couplings, so enhanced
contributions to the Yukawa couplings generally depend on
ratios of dimensionful soft SUSY-breaking parameters.   In addition, 
the size of supersymmetric particle effects is limited by the value of
the   heavy Higgs boson mass $m_A$.  Since the size of $m_A$ controls the
rate of decoupling, the possible size of deviations in the $b$ quark
Yukawa coupling falls off rapidly for $m_A$ above 3~TeV.  However,
the highest $A$ and $H$ masses here are already above the expected HL-LHC
search reach.

Given that $m_A$ is large enough, the sensitivity to
the gluino, the Higgsino,  and the $b$ squarks extend to masses 
well above 5~TeV.   In particular, the conditions for observable
contributions to the $b$ Yukawa coupling can be easily met in models
with  very large masses for the gluino and other gauginos.   This
conclusion  follows from the specific physics of heavy particle
decoupling that we have explained in Section II.

Comparison of the results of our exploration of a variety of BSM
models reinforces the idea that each such model predicts its own
characteristic pattern of deviations from the Standard Model in the
Higgs boson coupling.   The complete pattern of
deviations over the full range of Higgs boson couplings potentially
contains a large amount of information about physics beyond the
Standard Model.  This must be 
a central target of collider physics measurements after the era of the LHC.

\begin{acknowledgments}
 
 We are grateful to Howard Haber and Sven Heinemeyer for their advice
 concerning this project. The work of KM, DGEW, and MF-R
 was supported in part by the grant NSF
OIA-2033382.   The work of MEP was partially supported by the US Department
of Energy under contract DE–AC02–76SF00515.  
\end{acknowledgments}

\section{Data Availability}

The Mathematica notebook used for the calculations in
this paper is available at \cite{Maayergi2026}. The SLHA files generated from this scan are also available by contacting the authors.

\bibliography{Bibliography2}

\end{document}